\documentstyle[aps,prb,multicol,epsf]{revtex}
\begin{document}
\title{
Effects of Strong Correlation and Randomness in the Vicinity of
the Mott Transition in the Quasi-One-Dimensional Hubbard Model
}

\author{Satoshi Fujimoto}
\address{
Department of Physics,
Kyoto University, Kyoto 606, Japan
}

\date{\today}

\maketitle
\begin{abstract}
We study the strong correlation effects in the vicinity of 
the Mott metal-insulator transition
using coupled clean or disordered Hubbard chains with a infinitely 
large coordinate number $D_{\perp}\rightarrow\infty$
in the direction perpendicular to the chains and
with a long-range transverse hopping.
Strong electron correlation effects are treated partially non-perturbatively
with the use of the exact results for the 1D Hubbard model. 
In the case of clean systems, the thermodynamic and transport quantities 
which characterize the Mott transition 
from the Fermi liquid state,
such as the specific heat coefficient, the Drude weight, and 
the compressibility,
are obtained as functions of hole-doping $\delta=1-n$ ($n$, electron density)
by the systematic expansion in terms of
the inverse of the transverse hopping range $l_{\perp}$.
We find that the $\delta$-dependence of these quantities shows
non-universal behaviors with exponents depending on the strength of 
electron-electron interaction.
In the presence of disorder, it is shown
that the frequency-dependence of the dynamical conductivity 
obeys the Mott's law,
implying the possibility of {\it Mott glass state} 
for $\delta \rightarrow 0$,
provided that there exists a finite range interaction.
\end{abstract}

\section{Introduction}
The Mott-Hubbard(MH) metal-insulator transition  
is a long-standing important issue 
in strongly correlated electron systems.\cite{mott,hub,br,tokura}
There has been much progress in the theoretical study of 
the MH transition in the one-dimensional(1D) 
Hubbard model\cite{lieb,ky,sch,com} and 
the infinite-dimensional Hubbard model.\cite{kot,geo} 
However, our understanding for the MH transition
in the spatial dimensions between one and infinity 
is still far from complete.
In the metallic state near the MH transition, 
it is expected that the tendency toward 
the MH transition may appear in thermodynamic and transport
properties such as specific heat coefficient, compressibility, and
Drude weight in the metallic state, as the electron density approaches
to the half-filling, 
or the strength of electron correlation approaches to
a critical value $U_c$.
However, in general, the concrete computation of these quantities
in the vicinity of the MH transition is not tractable.
It is difficult to go beyond mean field theories such as
the Gutzwiller-type approximation 
in dimensions larger than one.\cite{br}
The purpose of this paper is to propose and solve the model
in which an advanced treatment of electron correlation
beyond conventional mean field theories is possible
in dimensions larger than one.

Here, we consider the coupled Hubbard chains 
with a large coordinate number $2D_{\perp}$
in the direction perpendicular to the chains, 
and with a long-range transverse hopping. 
We investigate how the sign of the MH transition
manifests in highly anisotropic Fermi liquids,
as the electron density $n$ approaches the half-filling.
Our strategy is as follows.
We expand the partition function in terms of the transverse hopping.
Then, we can deal with the intra-chain electron-electron scattering processes
exactly exploiting the Bethe ansatz exact solution
for the 1D Hubbard model.
The inter-chain electron-electron scattering processes are
treated perturbatively.
This partially exact treatment of electron-electron interaction
enable us to study strong correlation effects in the vicinity
of the MH transition beyond simple mean field theories.

We consider the following model Hamiltonian,
\begin{eqnarray}
H&=&-t\sum_{\sigma i,j}c^{\dagger}_{\sigma i,j}c_{\sigma i+1,j}+h.c.
+U\sum_{i,j} c^{\dagger}_{\uparrow i,j}c_{\uparrow i,j}
c^{\dagger}_{\downarrow i,j}c_{\downarrow i,j},  \nonumber \\
&&-\sum_{\sigma j,j'}t_{\perp jj'}c^{\dagger}_{\sigma i,j}c_{\sigma i,j'}.
\label{hub}
\end{eqnarray}
Here we introduce the long-range transverse hopping with a hopping range
$l_{\perp}$,
\begin{equation}
t_{\perp}(k_{\perp})=\frac{t_{\perp}}{l_{\perp}\sqrt{D_{\perp}}}
\sum_{n=-l_{\perp}}^{l_{\perp}}\sum_{j=1}^{D_{\perp}}\cos nk_{\perp j}.
\end{equation}
Several years ago, Boies {\it et al.} studied the two-dimensional
version of the above model in the context of 
dimensional crossover, and pointed out
that in the limit of infinitely long-range transverse hopping 
$l_{\perp} \rightarrow\infty$, the mean field solution of 
the Fermi liquid state is exact, and that the system is reduced 
to the non-interacting fermion system with a renormalized mass.\cite{boi}
Here we consider the expansion in $1/l_{\perp}$
around this non-interacting limit.
As will be shown in this paper,
in order to make the expansion in terms of $1/l_{\perp}$ 
controllable, we need to introduce a sufficiently large 
coordinate number in the transverse direction $2D_{\perp}$.
The critical dimension of $D_{\perp}$ will be discussed later.
For convenience, we consider the infinitely large $D_{\perp}$
which simplifies perturbative calculations in $1/l_{\perp}$.\cite{infd}
The $1/l_{\perp}$-corrections generate multi-particle interactions,
which give rise various interesting electron correlation effects.
Here, we consider the {\it paramagnetic} Mott transition,
since any magnetic phase transitions may conceal 
generic effects of electron correlation which drive
the system to the Mott insulator.
 
Another purpose of this paper is to investigate effects of disorder
in the vicinity of the Mott transition.
Very recently, it is proposed by Orignac {\it et al.} that
in the presence of both disorder and umklapp scattering due to
electron correlation, a novel state called {\it Mott glass} 
may realize.\cite{ori}
In this state, a single-particle excitation has a charge gap,
whereas particle-hole excitation is gapless, and
the dynamical conductivity shows the Mott's law, 
$\sigma(\omega)\sim\omega^2$ for small $\omega$. 
Orignac {\it et al.} used a 1D interacting spinless fermion model with disorder
to discuss this possibility.
However, as is well-known, the excitation gap of the 1D spinless 
fermion model at half-filling is destroyed by infinitesimally small
disorder.\cite{shan}
To avoid this difficulty,
they add the mass term to the model to stabilize the gapped state.
In contrast to the spinless fermion model, 
the Mott-Hubbard gap of the 1D Hubbard model
is stable against sufficiently small disorder.\cite{fujikawa,mori,hatsu}
Thus introducing random potential to our model, eq.(\ref{hub}), 
we can investigate more directly whether the Mott glass
is realized or not by disorder.
We would like to address this issue in this paper.

The organization of this paper is as follows.
In \S 2, we introduce the effective action which is the basis
of our approach, and discuss the case of the infinitely long range 
transverse hopping, where the system is reduced to the non-interacting
fermions with an enhanced mass.
In \S 3, we consider $1/l_{\perp}$-corrections to 
the results obtained in \S 2.
We see that the multi-particle interactions 
generated by the $1/l_{\perp}$-corrections 
change the doping-dependence of the thermodynamic and transport properties
which characterize the MH transition.
In \S 4, we discuss effects of disorder in the vicinity of
the Mott transition, and a possibility of the Mott glass state.
We give a summary in \S 5.

\section{Effective Action and the Fermi Liquid State
in the Limit of Infinitely Long-Range Transverse Hopping}

\subsection{Effective action}

In this subsection, we give the effective action of the model (\ref{hub})
which is suitable for the following argument.
In order to exploit the exact results of the 1D Hubbard model,
we expand the action in terms of the transverse hopping.
This approach was taken before
in the study of dimensional crossover between the Luttinger liquid
and the Fermi liquid.\cite{wen,boi}
Here we just outline the basic idea.
We introduce the Grassmann variable $\psi_{\sigma i,j}$,
applying the Hubbard-Stratonovich transformation,\cite{boi}
\begin{eqnarray}
&&t_{\perp jj'}c^{\dagger}_{\sigma i,j}c_{\sigma i,j'}
\rightarrow  \nonumber \\
&&t_{\perp jj'}c^{\dagger}_{\sigma i,j}\psi_{\sigma i,j'}
+t_{\perp jj'}\psi^{\dagger}_{\sigma i,j}c_{\sigma i,j'} 
+t_{\perp jj'}\psi^{\dagger}_{\sigma i,j}\psi_{\sigma i,j'}.
\end{eqnarray}
Then averaging over $c_{\sigma i,j}$, $c^{\dagger}_{\sigma i,j}$
with respect to the action for the 1D Hubbard model,
we can write the partition function in the following form,
\begin{equation}
Z=Z_{1D}\int {\cal D}\psi {\cal D}\psi^{\dagger}
e^{-S(\psi,\psi^{\dagger})},
\end{equation}
where $Z_{1D}$ is the partition function of the 1D Hubbad model,
and the effective action (free energy functional) 
$S(\psi,\psi^{\dagger})$ is given by 
\begin{eqnarray}
&&S(\psi,\psi^{\dagger})=S_1+\sum_{n=2}^{\infty}S_n, \\
&&S_1= \int {\rm d}x_1{\rm d}x_2\sum_{k_{\perp}}
\psi^{\dagger}_{\sigma k_{\perp}}(x_1)[t_{\perp}(k_{\perp})\delta(x_1-x_2)
-t_{\perp}^2(k_{\perp})G^{(1)}(x_1-x_2)]
\psi_{\sigma k_{\perp}}(x_2), \\
&&S_n=\frac{1}{n!N_{\perp}^{n-1}}\sum_{k_{\perp}}\int 
{\rm d}x_1...{\rm d}x_{2n}
T_{\perp}(k_{\perp 1},...,k_{\perp 2n}) 
G_c^{(n)}(x_1,x_2,...,x_{2n})
\psi^{\dagger}_{\sigma_1 k_{\perp 1}}(x_1)...\psi_{\sigma_{2n}k_{\perp 2n}}
(x_{2n}), \label{s2} \\
&&T_{\perp}(k_{\perp 1},...,k_{\perp 2n})=\prod_j^{2n}
t_{\perp}(k_{\perp j})\delta_{\sum_j^{2n}k_{\perp j},0}.
\end{eqnarray}
Here $x_i=(x_{\parallel}, t)$, 
$G^{(1)}(x_1-x_2)$ is the single-particle Green's function
of the 1D Hubbard model, and $G_c^{(n)}$ is the connected
$2n$-body correlation functions of the 1D Hubbard model.
$N_{\perp}$ is the total number of the Hubbard chains.
In this expression, all effects of electron correlations 
are incorporated into
correlation functions of the 1D Hubbard model.
Then we can utilize the exact result of the 1D Hubbard model.

\subsection{The case of infinitely long-range transverse hopping}

We, first, consider the limit of infinitely long-range transverse hopping,
$l_{\perp}\rightarrow \infty$.
As was pointed out by Boies {\it et al.},
in this limit, $S_{n\geq 2}$ vanish.
This is due to the fact that 
for $l_{\perp}\rightarrow \infty$,
$t_{\perp}(k_{\perp})\rightarrow (t_{\perp}/\sqrt{D_{\perp}})
\sum_{j=1}^{D_{\perp}}\delta_{k_{\perp j},0}$, and thus
the factor $1/N_{\perp}^{n-1}=1/l_{\perp}^{(n-1)D_{\perp}}$ in eq.(\ref{s2})
remains even after the summation over $k_{\perp}$.
As a result, the action is given by $S_1$, i.e.
the non-interacting fermion with the renormalized mass. 
The single-particle Green's function in this limit is given by,
\begin{equation}
G(k,\mbox{\boldmath $k$}_{\perp},\varepsilon)=
\frac{G^{(1)}(k,\varepsilon)}
{1-t_{\perp}(\mbox{\boldmath $k$}_{\perp})G^{(1)}(k,\varepsilon)}.
\label{sg1}
\end{equation}
Although the 1D Hubbard model is exactly solvable,
the exact expression for correlation functions
is known only for 
the low-energy Luttinger liquid state.\cite{hal,kawa,fk}
For simplicity, we use the following approximated 
expression for the Green's function 
of the Luttinger liquid in the momentum space,\cite{wen}
\begin{equation}
G_{L(R)}^{(1)}(k,\varepsilon)\sim 
\frac{E_0^{-\theta_c}(v_c^2k^2-\varepsilon^2)^{\theta_c/2}}
{\sqrt{(v_sk\pm\varepsilon)(v_ck\pm\varepsilon)}},
\end{equation}
where $v_c$ and $v_s$ are the velocities of holon and spinon, 
respectively, and 
$\theta_c=\frac{1}{4}(K_c+\frac{1}{K_c}-2)$ with 
$K_c$, the Luttinger liquid parameter. 
$E_0$ is an ultra-violet cutoff energy for the Luttinger liquid state.
Substituting this into eq.(\ref{sg1}),
one finds that away from half-filling, 
the quasiparticle pole of eq.(\ref{sg1}) exists, and thus
the system is the Fermi liquid. 
In this state, the Fermi surface is shifted by,
\begin{equation}
\Delta k_F=\biggl(\frac{v_c^{\theta_c}\vert t_{\perp}(k_{\perp})\vert}
{\sqrt{v_sv_c}E_0^{\theta_c}}\biggr)^{\frac{1}{1-\theta_c}}
\mbox{sgn}(t_{\perp}(k_{\perp})), \label{fermis}
\end{equation}
and the single-particle energy dispersion near the Fermi surface is
given by,  
\begin{equation}
E_k\sim 2(1-\theta_c)\frac{v_cv_s}{v_c+v_s}\vert k\vert
\equiv \tilde{v}\vert k\vert.
\end{equation}
We have the finite quasiparticle weight,
\begin{equation}
z_{k_F}=\frac{2t_{\perp}(k_{\perp F})E_0^{-2\theta_c}
(v_c\Delta k_F)^{2\theta_c}}{(v_c+v_s)\Delta k_F}.
\end{equation}
It is noted that the Fermi surface shift eq.(\ref{fermis})
is non-zero only at $k_{\perp j}=0$, and thus
the change of the Fermi surface volume due to this shift
vanishes for infinitely large $D_{\perp}$.
Namely, $\Delta V_{FS}/V_{FS}
=\Delta k_{F}/k_{F}(2\pi)^{D_{\perp}}\rightarrow 0$
in the limit of $D_{\perp}\rightarrow \infty$.
This is compatible with the Luttinger's theorem.

In this special limit, the effective action is given by,
\begin{equation}
S_{mf}=\sum_{k_{\perp}}\int {\rm d}x_{\parallel} {\rm d}t 
\psi^{\dagger}_{\sigma k_{\perp}}(x_{\parallel},t)z_{k_F}^{-1}
(\partial_t-\tilde{v}\partial_{x_{\parallel}})
\psi_{\sigma k_{\perp}}(x_{\parallel},t). \label{meanef}
\end{equation}
This describes the Fermi liquid mean field solution (RPA solution)
of the coupled chains.

Using the these expressions, we can investigate
correlation effects in the vicinity of MH transition.
According to the Bethe ansatz exact solution,
near the half-filling, 
$v_c\sim \delta$ ($\delta=1-n$, $n$, electron density).
$E_0$ is roughly given by $v_c(\pi-2k_F)/2\sim \delta^2$.
It should be noted that the above expressions are applicable
provided that the shift of the Fermi surface $\Delta k_F$ is
smaller than $\pi\delta/2$.
If this condition is not satisfied, we can not use the expression
of the correlation function of the Luttinger liquid.
Therefore our following argument is restricted to the case of
$\delta > ct_{\perp}^{2/3}$, where $c$ is a constant of order unity.
However, for sufficiently small $t_{\perp}$, we can approach
the vicinity of MH transition point, where 
strong correlation effects
characterizing the metal-insulator transition manifest.
Near the half-filling but for $\delta > ct_{\perp}^{2/3}$, 
the quasiparticle weight is given by,
\begin{equation}
z_k\sim\delta^{(1-4\theta_c)/2(1-\theta_c)}
t_{\perp}^{\theta_c/(1-\theta_c)}. \label{qw1}
\end{equation}
Thus the quasiparticle weight is much suppressed,
and decreases toward zero, as the electron density approaches 
the half-filling. 

Now we investigate thermodynamic and transport
properties in the vicinity of MH transition.
Thermodynamic quantities are given by the sum of the contribution
from the 1D Hubbard model and the Fermi liquid state;
i.e. $Q=Q_{1D}+Q_{FL}$.
In this paper, we concentrate on $Q_{FL}$ 
in order to see the correlation effects
in the Fermi liquid state near the Mott transition.
We can easily calculate the $\delta$-dependence of the 
Drude weight $D_0$, the specific heat coefficient $\gamma_0$, and
the compressibility $\kappa_0$ of the Fermi liquid part
in this non-interacting limit,
\begin{equation}
D_0=\sum_k v_k^2\delta(\mu-E_k)\sim \delta,
\end{equation}
\begin{equation}
\gamma_0=\sum_k\delta(\mu-E_k)\sim \delta^{-1},
\end{equation}
\begin{equation}
\kappa_{c0}=\sum_kz_k^2\delta(\mu-E_k)
\sim \delta^{-\frac{3\theta_c}{1-\theta_c}}.
\end{equation}

The behaviors of the Drude weight and the specific heat coefficient
are similar to those of the 1D Hubbard model.
It is noted that $D_0\gamma_0$ is independent of $\delta$.
This is easily understood since there are no vertex corrections
in this non-interacting limit.

For the 1D Hubbard model, $0\leq \theta_c\leq 1/8$.
Thus $\kappa_{c0}$ shows a divergent behavior,
as the electron filling approaches the half-filling.
This is also due to the lack of vertex corrections.
However, in contrast to the 1D Hubbard model,
where the compressibility behaves like $\kappa\sim \delta^{-1}$,
the exponent of $\kappa_{c0}$ is non-universal depending on
the strength of interaction.\cite{com} 

The single-particle density of states of electrons 
at the Fermi surface also shows a remarkable divergent behavior,
\begin{equation}
\rho(0)=\sum_kz_k\delta(\mu-E_k)\sim 
\delta^{-\frac{1+2\theta_c}{2(1-\theta_c)}}.
\label{dos}
\end{equation}
This result makes a sharp contrast to the MH transition 
in the infinite-dimensional Hubbard model, where 
the density of state remains finite.\cite{kot,geo,mul1}
The divergent behavior of the density of states 
is also similar to the MH transition of the 1D Hubbard model.
However the exponent of eq.(\ref{dos}) 
is much modulated by electron correlation.



\section{Finite-Hopping Range Corrections}

In this section, we consider the $1/l_{\perp}$-corrections
which generate $2n$-body interaction as shown in Fig. 1.
Then the model can not be solved exactly anymore.
What we can do is to perform a perturbative expansion
in terms of $1/l_{\perp}$.
First of all, we give a scaling argument and justify
our perturbative approach in the next subsection.

\subsection{Renormalization group argument}

We consider correlation effects beyond 
the mean field solution given by $S_{mf}$, eq.(\ref{meanef}).
We apply the scaling transformation which keeps 
$S_1=S_{mf}$, eq.(\ref{meanef}), invariant,\cite{shan}
\begin{eqnarray}
\vec{x} \rightarrow \lambda \vec{x}, &\qquad&
t \rightarrow \lambda t, \nonumber \\
\psi_{\sigma}(\vec{x},t) &\rightarrow& 
\lambda^{\alpha}\psi_{\sigma}(\lambda\vec{x},\lambda t).
\end{eqnarray}
Then, the scaling dimension of $\psi_{\sigma}(\vec{x},t)$
reads $\alpha=\frac{D_{\perp}+1}{2}$.
Now we consider the scaling equation for
the effective coupling of $S_n$ ($n\geq 2$),
$g_n\equiv (t_{\perp}/l_{\perp})^{2n}$.
Applying the perturbative renormalization group argument,
we obtain the scaling equation,
\begin{equation}
\frac{{\rm d} g_n}{{\rm d} \ln \Lambda}=[3n-(n-1)D_{\perp}-\gamma_n]g_n.
\label{sca}
\end{equation}
Here $\Lambda$ is the dimensionless scaling parameter which goes to
infinity in the low-energy limit, and $\gamma_n\geq 0$ is 
the scaling dimension of $G_c^{(n)}(x_1,x_2,...,x_{2n})$.
Although we know the value of $\gamma_n$ for the 1D Hubbard model 
from the conformal field thoery argument,\cite{kawa,fk}
$G_c^{(n)}(x_1,x_2,...,x_{2n})$ in $S_{n\geq 2}$ may show a deviation
from the critical power law behavior in the long-distance limit  
owing to the deformation of the Fermi surface discussed above.
In the conventional Ginzburg-Landau arguments 
of coupled chains,\cite{sip} it is often assumed that 
the coefficient of the fourth order term is a non-zero constant. 
This means $\gamma_n=0$.
However the concrete value of $\gamma_n$ does not matter 
in the following argument.
We just assume it is finite.
Then we can easily see that for $D_{\perp}> (3n-\gamma_n)/(n-1)$, 
$S_n$ with $n\geq 2$ is irrelevant
in the low-energy state, and can be treated perturbatively. 
The existence of the critical dimension for the irrelevance
of $S_{n\geq 2}$ means nothing more than the stability of
the RPA like mean field solution in the large dimensional system
with the long-range inter-chain interaction,
which is consistent with the common sense that
a mean field solution is more stabilized 
in the presence of long-range interaction in spatial dimension larger
than an upper critical dimension.
In this paper, we consider the infinitely large $D_{\perp}$
which simplifies the perturbative calculation
as well as ensures the controllable expansion.
In this limit, the self-energy does not depend on
the transverse momentum $k_{\perp}$,
as in the case of the infinitely large dimensional Hubbard model,
where the self-energy is a completely local quantity.\cite{mez,mul2}

In the following, we investigate strong correlation effects
by performing the perturbative calculation in $1/l_{\perp}$
Before doing that,  we need to care about any possibility
of spontaneous symmetry breaking such as magnetic transition
which may be brought about by the interaction between quasiparticles.
However, in this paper,
we assume that no spontaneous symmetry breaking occurs 
before the MH transition,
since we are concerned with correlation effects in the Fermi liquid
state near the paramagnetic MH transition point.

Before carrying out the perturbative calculation,
we would like to make a comment about the Luttinger's theorem
which holds for the case of finite $l_{\perp}$.
The change of the Fermi surface volume due to the Fermi surface shift, 
eq.(\ref{fermis}), is given by
$\Delta V_{FS}=2\sum_{k_{\perp}}\Delta k_{F}$.
In the case of $D_{\perp}\rightarrow\infty$,
we have,
\begin{equation}
\Delta V_{FS}={\rm const.}\int^{\infty}_{-\infty} 
dt\vert t\vert^{\frac{1}{1-\theta_c}}
{\rm sgn}(t)\rho_{\perp}(t)=0,
\end{equation}
where 
\begin{equation}
\rho_{\perp}(t)\equiv 
\sum_{k_{\perp}}\delta(t-t_{\perp}(k_{\perp}))=
\frac{\sqrt{l_{\perp}}}{\sqrt{\pi}t_{\perp}}
e^{-\frac{l_{\perp}t^2}{t_{\perp}^2}}.
\end{equation}
Thus the Fermi surface volume is preserved.
This is consistent with the Luttinger's theorem.

\subsection{Perturbative calculation of the self-energy}

We evaluate the $1/l_{\perp}$-corrections to the self-energy.
We neglect the diagrams which just give a chemical 
potential shift, as shown in Fig. 2(a),(b),(c).
The diagram of order $O(g_2^2)$ (Fig. 2(d)) is most important
for the mass enhancement factor.
The self-energy correction from this diagram
is given by,
\begin{equation}
\Sigma_k^{(d)}(\varepsilon_n)\sim 
g_2^2\delta^{-\frac{1+8\theta_c}{2(1-\theta_c)}}{\rm i}\varepsilon_n
+...,
\label{selfc1}
\end{equation}
The next order diagram which contributes to the mass enhancement factor 
is of order $O(g_2^3)$(Fig. 2(e)),
since $g_3^2\ll g_2^3$ as seen from eq.(\ref{sca}).
It is evaluated as,
\begin{equation}
\Sigma_k^{(e)}(\varepsilon_n)\sim g_2^3
\delta^{-\frac{1+14\theta_c}{2(1-\theta_c)}}{\rm i}\varepsilon_n
+....
\label{selfc2}
\end{equation}
Note that $\Sigma^{(e)}$ shows more singular $\delta$-dependence
than $\Sigma^{(d)}$.
Higher order corrections also give more singular $\delta$-dependence.
Thus we need to be careful with the applicability of
our perturbative calculation. 
According to eq.(\ref{sca}),
$g_2\sim 1/\Lambda^{D_{\perp}-6}$.
Then, the ratio of eq.(\ref{selfc2}) to eq.(\ref{selfc1}) is 
sufficiently small provided that $\Lambda \gg 
(\delta^{-3\theta_c/(1-\theta_c)})^{1/(D_{\perp}-6)}$.
The applicability of the result of order $O(g_2^2)$, (\ref{selfc1}),
is restricted to this low-energy scale.
In other words, as the electron density approaches the half-filling
$\delta\rightarrow 0$, the energy range where eq.(\ref{selfc1})
can be used becomes very narrower such as 
$\Delta E\ll E_F(\delta^{3\theta_c/(1-\theta_c)})^{1/(D_{\perp}-6)}$.
In the following, we restrict our argument to this energy range.

\subsection{Specific heat coefficient}

Using eq.(\ref{selfc1})
we have the mass renormalization factor,
\begin{equation}
z_{k}^{*}=\frac{1}{1-\frac{\partial \Sigma_k(E_k^{*})}{\partial \varepsilon}}
\sim \delta^{\frac{1+8\theta_c}{2(1-\theta_c)}}.
\label{mef}
\end{equation}
The quasiparticle weight is much more suppressed compared to
eq.(\ref{qw1}).
In contrast, the quasiparticle damping 
$z_k^{*}{\rm Im}\Sigma_k(\varepsilon)$
is still infinitesimally small in the low-energy state.
Thus the Fermi liquid theory is applicable even in the vicinity of 
the MH transition.
We obtain the $\delta$-dependence of 
the specific heat coefficient,\cite{lutt}
\begin{equation}
\gamma=\frac{\pi^2}{3}\sum_k\delta(\mu -E_k^{*})
\sim \delta^{-\frac{1+5\theta_c}{1-\theta_c}}.
\label{shc2}
\end{equation}
The specific heat coefficient shows the divergent behavior
as $\delta$ decreases. The mass enhancement is larger
than that of the non-interacting limit obtained in \S 2. 
Note that the exponent of eq.(\ref{shc2}) depends on the strength
of interaction indicating the non-universal behavior.

\subsection{Single-particle density of states}

Up to the second order corrections in $g_2$,
the single-particle density of states at the Fermi surface 
is given by,
\begin{equation}
\rho(0)=\sum_k z_k^{*}\delta(\mu-E_k^{*})\sim 
\delta^{-\frac{1+2\theta_c}{2(1-\theta_c)}}.
\end{equation}
The result is the same as that obtained in \S 2(eq.(\ref{dos})).
Higher-order corrections does not affect this result,
as expected from the conventional Fermi liquid theory.  

\subsection{Drude weight}

In the computation of the Drude weight,
we take into account the effect of the velocity renormalization,
$v_k^{*}=z_k^{*}\partial_k(E_k+{\rm Re}\Sigma_k(0))$ 
due to the self-energy correction, eq.(\ref{selfc1}), as well as
the vertex corrections to electric currents which is nothing but
the back flow effect,\cite{noz}
\begin{equation}
J_k=v_k^{*}+\sum_{k'}z_k^{*}z_{k'}^{*}\Gamma(k,k')\delta(\mu-E_{k'}^{*})
v_{k'}^{*}.
\end{equation}
Here $\Gamma(k,k')$ is the interaction between the quasiparticles
with momenta $k$ and $k'$.
When we calculate the current vertex corrections, 
we generally need to pay attention to the charge conservation law.
However the charge conservation is not trivial in our system.
The non-perturbed state obtained in \S 2 
is composed of the Fermi liquid part
and the Luttinger liquid part.
As a result, the weight of the Fermi liquid part may be 
transfered to the Luttinger liquid part
by the interaction introduced in this section.
The charge conservation law is not expressed in the closed form
only in the Fermi liquid part.
Thus, here, 
we just evaluate the vertex corrections up to the second order 
in $g_2$, of which the diagrams are shown in Fig. 3.
Note that each line in Fig.3 is the renormalized single-particle
Green's function with the quasiparticle weight $z_k^{*}$, eq. (\ref{mef}).
This renormalization effect must be taken into account, 
because the use of the bare propagator eq.(\ref{sg1})
for the calculation of the diagrams in Fig. 3
leads to unphysical enhancement of current vertex corrections
for small $\delta$. 
This treatment may be justified by the Ward-Takahashi identity,
though we do not discuss this point furthermore because of 
the reason mentioned above.
Then we have the Drude weight,
\begin{equation}
D=e^2\sum_kv_k^{*}J_k\delta(\mu -E_k^{*})\sim 
\delta^{\frac{1+5\theta_c}{1-\theta_c}}.
\end{equation}
It is noted that 
the relation, $D\gamma=\delta$-independent
still holds {\it approximately}.
This is because that the corrections due to the back flow effect
is sub-leading in this systems in the vicinity of the MH transition
and thus the leading behavior of the Drude weight
is determined by the velocity of quasiparticles.

\subsection{Compressibility} 

Here we will see that
the effect of the interaction between quasiparticles
affects the $\delta$-dependence of the compressibility
drastically.
Using the dressed single-particle Green's function
with the mass renormalization factor $z_k^{*}$,
as in the case of the Drude weight,
we compute the lowest order diagram for the compressibility as shown 
in Fig. 4(a),
\begin{eqnarray}
\kappa_c \sim \delta^{\frac{3\theta_c}{1-\theta_c}}.
\end{eqnarray}
Higher order diagrams shown in Fig. 4(b)
give sub-dominant contributions for small $\delta$.

In contrast to the result in \S 2,
as $\delta\rightarrow 0$, $\kappa_c$ decreases toward zero,
though we can not reach the half-filling point, $\delta=0$, because of
the restriction $ct_{\perp}^{2/3} < \delta$. 
In the conventional Fermi liquid theory,
it is expected that vertex corrections due to 
electron-electron interaction suppress the compressibility.
However, here, the lowest order bubble diagram Fig. 4(a)
gives the strong suppression of the compressibility.
This implies that the {\it usual} Fermi liquid effect of vertex corrections 
may be incorporated effectively
into the mass renormalization factor $z_k^{*}$ in our model.
The drastic change of $\delta$-dependence 
due to $1/l_{\perp}$-corrections
means that $S_{n\geq 2}$ terms, eq.(\ref{s2}), 
are in a sense {\it dangerously irrelevant} for $D_{\perp}>6$.

\section{Effects of Disorder: a Possibility of Mott Glass}

In this section, we study the effects of random potential
in the vicinity of the Mott transition.
We are mainly concerned with a possible realization of the Mott glass
state proposed by Orignac {et al}.\cite{ori}
The Mott glass state is characterized by vanishing compressibility and
the dynamical conductivity which obeys the Mott's law, namely,
$\sigma(\omega)\sim \omega^2$ for small $\omega$.
In this state, although a single-particle exciation has a charge gap,
particle-hole excitation is gapless showing
the characteristic behavior of Anderson localization.
Thus the Mott glass state is regarded as an intermediate state
between Mott insulator and Anderson localization. 
This state is similar to Anderson localized state in the
presence of long-range Coulomb interaction 
in the spatial dimension larger than 2.\cite{efros}
However the Mott glass state is realized for the case of 
finite-range interaction and even for 1D systems.
To see a possible realization of this state,
we add the disorder potential term to the Hamiltonian:
\begin{equation}
H_{dis}=\sum_{i,j}\xi_{i,j}c^{\dagger}_{\sigma i,j}c_{\sigma i,j}.
\end{equation}
Here $\xi_{i,j}$ is a random potential.
In the presence of disorder, we can derive the effective action 
in a simalr way to that applied for clean systems.
Using the Hubbard-Stratonovich transformation,
\begin{eqnarray}
A_{j,j'}&\equiv& t_{\perp jj'}+\delta_{jj'}\xi_{ij}, \nonumber \\
A_{jj'}c^{\dagger}_{\sigma i,j}c_{\sigma i,j'}
&\rightarrow& A_{jj'}c^{\dagger}_{\sigma i,j}\psi_{\sigma i,j'}
+A_{jj'}\psi^{\dagger}_{\sigma i,j}c_{\sigma i,j'} 
+A_{jj'}\psi^{\dagger}_{\sigma i,j}\psi_{\sigma i,j'},
\end{eqnarray}
we decouple the disorder term as well as the transverse hopping term.
Then integrating out $c$, $c^{\dagger}$ fields, we have,
\begin{eqnarray}
&&S(\psi,\psi^{\dagger})=S_1+\sum_{n=2}^{\infty}S_n, \\
&&S_1= \int dx_1dx_2\sum_{k_{\perp}}
\psi^{\dagger}_{\sigma k_{\perp}}(x_1)[A(x_{\parallel},k_{\perp})
\nonumber \\
&&-A(x_{\parallel 1},k_{\perp})A(x_{\parallel 2},k_{\perp})G^{(1)}(x_1-x_2)]
\psi_{\sigma k_{\perp}}(x_2), \\
&&S_n=\frac{1}{n!N_{\perp}^{n-1}}\sum_{k_{\perp}}\int dx_1...dx_{2n}
B(x_{\parallel 1},...,x_{\parallel 2n};k_{\perp 1},...,k_{\perp 2n}) 
\nonumber \\
&& \qquad\qquad \times G_c^{(n)}(x_1,x_2,...,x_{2n})
\psi^{\dagger}_{\sigma_1 k_{\perp 1}}(x_1)...\psi_{\sigma_{2n}k_{\perp 2n}}
(x_{2n}), \label{disac}\\
&&B(x_{\parallel 1},...,x_{\parallel 2n};
k_{\perp 1},...,k_{\perp 2n})=\prod_j^{2n}
A(x_{\parallel j},k_{\perp j})\delta_{\sum_j^{2n}k_{\perp j},0}. 
\end{eqnarray}
Since our model has the infinite $D_{\perp}$, Anderson localization
in the transverse direction does not occur.
For this reason, we concentrate on effects of disorder only 
in the direction parallel to the chains,
and consider randomness
without spatial dependence in the transverse direction, namely,
\begin{eqnarray}
\xi_{i,j}&\equiv& \xi(x_{\parallel i}), \nonumber \\ 
\langle \xi(x_{\parallel i})\xi(x_{\parallel k})\rangle&=&
D_{\xi}\delta(x_{\parallel i}-x_{\parallel k}).
\end{eqnarray}
This assumption simplifies the following argument substantially.
Moreover we consider sufficiently small disorder to
stabilize the single-particle charge gap.
It is known that the Mott-Hubbard gap of the 1D Hubbard model 
is not collapsed by sufficiently small disorder.\cite{fujikawa,mori,hatsu}
In the following, we take into account only the backward scattering
due to random potential and neglect the forward scattering
which is not important for the realization of the Mott glass.
Using the renormalization group argument similar to that
developed in \S 3.1, we can easily show that
$S_{n\geq2}$ of eq.(\ref{disac}) are irrelevant in the low energy state.
Thus we neglect the disorder effects of $S_{n\geq2}$, and
take into account the lowest order corrections in $1/l_{\perp}$
as done in \S 3.
Then, we have the low energy effective action,
\begin{equation}
S_{eff}=\int dx d\omega \psi_{\sigma k_{\perp}\omega}^{\dagger}(x_{\parallel})
[z_{k_F}^{*-1}(\omega-v_F^{*}\partial_{x_{\parallel}})-\xi(x_{\parallel})]
\psi_{\sigma k_{\perp}\omega}(x_{\parallel})+\mbox{irrelevant interactions}.
\end{equation}
In the following, we neglect the irrelevant interactions of this action.
As a result, the model is equivalent to the 1D disordered 
non-interacting fermion system, 
which has been studied extensively so far.\cite{bere,abri}
However the important difference from the classical model is
that the effects of electron correlation are
incorporated into the mass renormalization factor $z_{k_F}^{*}$ and
the renormalized Fermi velocity $v_F^{*}$.
Following Berezinskii,\cite{bere} and Abrikosov and Ryzhkin,\cite{abri}
we can caluculate the dynamical conductivity.
The result is given by,
\begin{equation}
\sigma(\omega)\sim \frac{v_F^{* 4}\omega^2}{D_{\xi}^3z_{k_F}^{* 6}}
\biggl(\ln\biggl(\frac{D_{\xi}z_{k_F}^{* 2}}{\omega v_F^{*}}\biggr)\biggr)^2.
\end{equation}
This expression shows the Mott's law, 
indicating the Anderson localized state.
For small $\delta$, it behaves like,
\begin{equation}
\sigma(\omega)\sim  
\frac{\delta^{\frac{1-4\theta_c}{1-\theta_c}}\omega^2}
{D_{\xi}^3}\biggl(\ln\biggl(\frac{D_{\xi}z_{k_F}^{* 2}}{\omega v_F^{*}}
\biggr)\biggr)^2.
\end{equation}
For the 1D Hubbard model, $0\leq \theta_c \leq 1/8$.
Thus as $\delta \rightarrow 0$, it is expected that $\sigma(\omega)$ vanishes,
though we can not reach $\delta=0$ because of the reason mentioned 
in the previous sections.
In this case, the Mott glass state may not realize at half-filling. 
However in the presence of a finite range interaction, such as
next-nearest or next-next-nearest intearction,
it is possible to be $\theta_c=1/4$ for $\delta\rightarrow 0$.
It is noted that even for $\theta_c=1/4$, the single-particle
Green's function, eq.(\ref{sg1}) still has a quasiparticle pole.
Moreover the velocity of holon near the half-filling point
behaves like $v_c\sim \delta$ even in the presence of
a finite-range interaction provided that the low-energy
effective theory of the single chain is given by Sine-Gordon model.
Thus the argument presented in the previous sections is applicable.
Then, in this case, there is a possibility of the Mott glass state 
with $\sigma(\omega)\sim \omega^2$ for $\delta\rightarrow 0$.
This result is consistent with the physical intuition
that the presence of a finite-range interaction reduces
particle-hole excitation energy lower than 
a single-particle excitation energy, stabilizing 
the Mott glass state.
Since our approach breaks down just at the half-fillig,
the above result is not a direct evidence of the Mott glass state.
However it implies a strong possibility of its realization
at half-filling.

\section{Summary and Discussion}

We have studied strong correlation effects
of the Fermi liquid state near the MH transition 
in the quasi-one-dimensional Hubbard model.
We have treated the intra-chain electron correlation exactly
exploiting the exact results of the Bethe ansatz solution.
The systematic perturbative expansion in terms of the inverse
of the transverse hopping range has been used for the analysis of
the inter-chain electron correlation effects.
It has been shown that 
the precursor of the MH transition can be seen in the
$\delta$-dependence of the Drude weight, 
the specific heat coefficient, and the compressibility,
and that the exponents which depend on the strength of 
electron-electron interaction characterize
non-universal behaviors of the MH transition.
We have also studied effets of disorder on this strongly correlated
state. We otain the dynamical conductivity which obeys the Mott's law.
The result implies a possible realization of the Mott glass state
at half-filling.

In real systems, the situation with a large $l_{\perp}$
hardly realizes. We would like discuss about to what extend
our results are valid in systems with small $l_{\perp}$.
In principle, we can incorporate the finiteness of $l_{\perp}$
by the systematic expansion.
As shown in \S 3, if we take into account the second-order
self-energy corrections, we have the Fermi liquid state with
a strongly renormalized quasi-particle weight, and then
near half-filling, higher order corrections in $1/l_{\perp}$
are much suppressed. (see the calculation of the Drude weight and
the compressibility in \S 3.)
Thus it is expected that the physical picture of {\it weakly-interacting 
quasi-particles with a strongly renormalized mass} 
may be hold in the vicinity of the Mott transition
even for the case with small $l_{\perp}$,
though the exponents of physical quantities should be quantatively
changed from those obtained in \S 3.

As claimed in \S 2., the applicability of our results is restricted to
$\delta > ct_{\perp}^{2/3}$. 
Thus, unfortunately, we can not reach the half-filling point.
We need the information for off-critical behaviors of correlation
functions of the 1D Hubbard model to investigate the case of 
$0\leq \delta <ct_{\perp}^{2/3}$.
Recently, the substantial progress of the study of 
correlation functions of 1D massive integrable models
has been achieved.\cite{lu}
It may be possible to apply the new technique for 
the case of $0\leq \delta <ct_{\perp}^{2/3}$.
We would like to address this issue in the near future.

\section*{Acknowledgment}
The author would like to thank N. Kawakami and K. Yamada
for invaluable discussions.
This work was partly supported by a Grant-in-Aid from the Ministry
of Education, Science, Sports and Culture, Japan.


\newpage

\begin{figure}
\centerline{\epsfxsize=8cm \epsfbox{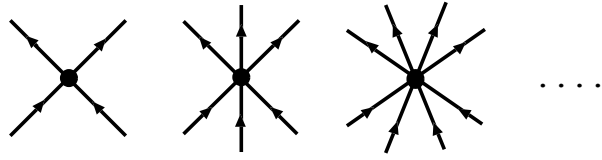}}
\caption{Some examples of multi-particle interaction.}
\end{figure}
\begin{figure}
\centerline{\epsfxsize=9.5cm \epsfbox{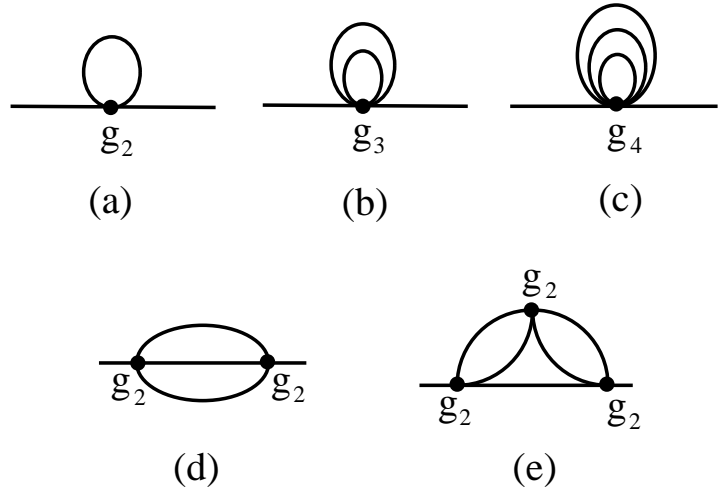}}
\caption{Diagrams for selfenergy corrections of order
(a) $O(g_2)$, (b) $O(g_3)$, (c) $O(g_4)$, (d) $O(g_2^2)$, 
(e) $O(g_2^3)$.}
\end{figure}
\begin{figure}
\centerline{\epsfxsize=6.8cm \epsfbox{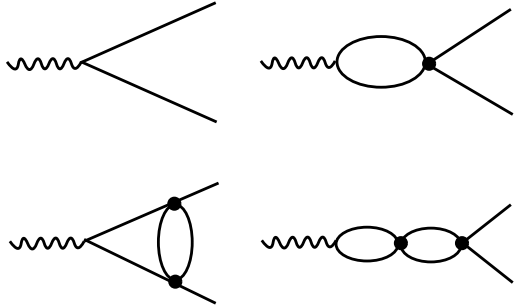}}
\caption{Diagrams for vertex corrections to the electric current 
up to the second order in $g_2$.}
\end{figure}
\begin{figure}
\centerline{\epsfxsize=12cm \epsfbox{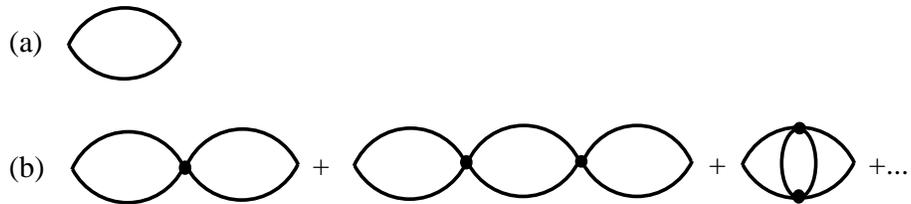}}
\caption{(a) Diagrams for compressibility without vertex corrections.
(b) Some examples of diagrams for compressibility 
with vertex corrections}
\end{figure}

\end{document}